\documentstyle[psfig]{mn}

\newif\ifAMStwofonts

\ifoldfss
  \ifCUPmtlplainloaded \else
    \NewTextAlphabet{textbfit} {cmbxti10} {}
    \NewTextAlphabet{textbfss} {cmssbx10} {}
    \NewMathAlphabet{mathbfit} {cmbxti10} {} 
    \NewMathAlphabet{mathbfss} {cmssbx10} {} 
  \fi
  \ifAMStwofonts
    \ifCUPmtlplainloaded \else
      \NewSymbolFont{upmath} {eurm10}
      \NewSymbolFont{AMSa} {msam10}
      \NewMathSymbol{\upi}     {0}{upmath}{19}
      \NewMathSymbol{\umu}     {0}{upmath}{16}
      \NewMathSymbol{\upartial}{0}{upmath}{40}
      \NewMathSymbol{\leqslant}{3}{AMSa}{36}
      \NewMathSymbol{\geqslant}{3}{AMSa}{3E}

      \let\leq=\leqslant 
       
    \fi
  \fi
\fi 

\ifnfssone
  \newmathalphabet{\mathit}
  \addtoversion{normal}{\mathit}{cmr}{m}{it}
  \addtoversion{bold}{\mathit}{cmr}{bx}{it}
  \newmathalphabet{\mathbfit} 
  \addtoversion{normal}{\mathbfit}{cmr}{bx}{it}
  \addtoversion{bold}{\mathbfit}{cmr}{bx}{it}
  \newmathalphabet{\mathbfss} 
  \addtoversion{normal}{\mathbfss}{cmss}{bx}{n}
  \addtoversion{bold}{\mathbfss}{cmss}{bx}{n}
  \ifAMStwofonts
    \ifCUPmtlplainloaded \else
      \UseAMStwoboldmath
      \makeatletter
      \new@mathgroup\upmath@group
      \define@mathgroup\mv@normal\upmath@group{eur}{m}{n}
      \define@mathgroup\mv@bold\upmath@group{eur}{b}{n}
      \edef\UPM{\hexnumber\upmath@group}
      \new@mathgroup\amsa@group
      \define@mathgroup\mv@normal\amsa@group{msa}{m}{n}
      \define@mathgroup\mv@bold\amsa@group{msa}{m}{n}
      \edef\AMSa{\hexnumber\amsa@group}
      \makeatother
      \mathchardef\upi="0\UPM19
      \mathchardef\umu="0\UPM16
      \mathchardef\upartial="0\UPM40
      \mathchardef\leqslant="3\AMSa36
      \mathchardef\geqslant="3\AMSa3E

      \let\leq=\leqslant 

    \fi
  \fi
\fi 

\ifnfsstwo
  \DeclareMathAlphabet{\mathbfit}{OT1}{cmr}{bx}{it}
  \SetMathAlphabet\mathbfit{bold}{OT1}{cmr}{bx}{it}
  \DeclareMathAlphabet{\mathbfss}{OT1}{cmss}{bx}{n}
  \SetMathAlphabet\mathbfss{bold}{OT1}{cmss}{bx}{n}
  \ifAMStwofonts
    \ifCUPmtlplainloaded \else
      \DeclareSymbolFont{UPM}{U}{eur}{m}{n}
      \SetSymbolFont{UPM}{bold}{U}{eur}{b}{n}
      \DeclareSymbolFont{AMSa}{U}{msa}{m}{n}
      \DeclareMathSymbol{\upi}{0}{UPM}{"19}
      \DeclareMathSymbol{\umu}{0}{UPM}{"16}
      \DeclareMathSymbol{\upartial}{0}{UPM}{"40}
      \DeclareMathSymbol{\leqslant}{3}{AMSa}{"36}
      \DeclareMathSymbol{\geqslant}{3}{AMSa}{"3E}

      \let\leq=\leqslant 

    \fi
  \fi
\fi 

\ifCUPmtlplainloaded \else
  \ifAMStwofonts \else 
    \def\upi{\pi}
    \def\umu{\mu}
    \def\upartial{\partial}
  \fi
\fi

\title{RXTE observation of NGC 6240: a search for 
the obscured active nucleus}
\author[Y. Ikebe et al.]
       {Yasushi Ikebe,$^1$ Karen Leighly,$^2$
	Yasuo Tanaka,$^{1,3}$ Takao Nakagawa,$^3$
    \newauthor 
	Yuichi Terashima,$^4$ and Stefanie Komossa$^1$ \\
  $^1$Max-Planck-Institut f\"{u}r extraterrestrische physik,
       Postfach 1603, D-85740, Garching, Germany\\
  $^2$Columbia Astrophysics Laboratory,
       550 West 120th Street, New York, NY 10027, USA\\
  $^3$Institute of Space and Astronautical Science, 
	Yoshinodai 3-1-1, Sagamihara, Kanagawa, 229-8510, Japan\\
  $^4$Laboratory for High Energy Astrophysics, NASA/GSFC, 
	Greenbelt, MD 20771, USA}

\pagerange{\pageref{firstpage}--\pageref{lastpage}}
\pubyear{2000}

\begin{document}

\maketitle

\label{firstpage}

\begin{abstract}
The wide-band energy spectrum of NGC~6240 over the range 0.5--200~keV
is investigated using the {\it RXTE} and {\it ASCA} data.
The {\it RXTE} data provide the spectrum beyond the {\it ASCA} range
(0.5--10 keV) with significant detection of signals up to 20~keV 
and the upper limits above 20~keV.
The spectrum above 10~keV is found to be very flat.
A strong iron-K emission line discovered in the previous {\it ASCA} observation
is also confirmed with the {\it RXTE} PCA.
These results provide further evidence for the dominance of a reflection
component, i.e. emission from cool material illuminated by an AGN.
By fitting the spectra obtained with {\it RXTE} and {\it ASCA} simultaneously,
we satisfactorily modeled the AGN spectrum
with a Compton reflection component and probably a
transmitted AGN component penetrating through a thick absorber.
The X-ray luminosity of the AGN is estimated to be in the range
$4 \times 10^{43} - 6 \times 10^{44}$ ergs/s in the range 2--10~keV,
which categorizes NGC~6240 among the most luminous Seyfert nuclei.
The ratio of the 2--10~keV X-ray luminosity to the infrared luminosity,
$L_{\rm X}{\rm (2-10~keV)}/L_{\rm IR}$, is 0.01 -- 0.1, which implies a
quite substantial, if not dominant, contribution of AGN to the infrared
luminosity.
\end{abstract}


\section{Introduction}
The IRAS survey (Neugebauer et al. 1984) discovered many
ultraluminous infrared galaxies (ULIRGs)
that emit the bulk of their energy in infrared (IR) photons.
Since their bolometric luminosity and the number density
are as high as those of quasars, 
ULIRGs are among the most energetic objects in the universe.
The most fundamental problem yet to be solved is the energy source
of the extremely intense infrared emission.

NGC~6240, a gravitationally interacting system with a complex
optical morphology (Fosbury \& Wall 1979; Fried \& Schulz 1983),
is a very interesting example of ULIRG.
Its bolometric luminosity reaches $2.4\times10^{12}$ $L_{\odot}$
(Weight, Joseph, \& Meikle 1984;
z=0.0245 and $H_0=50$ km s$^{-1}$Mpc$^{-1}$ are assumed).
NGC~6240 is outstanding in several respects.
Its H$_2$ $1\rightarrow0S(1)$ at 2.121$\mu$m 
and [FeII] 1.644$\mu$m line luminosities
and the ratio of H$_2$ to bolometric luminosities are the
largest currently known (e.g., van der Werf et al. 1993).
Further, its stellar velocity dispersion of 360 km/s
is among the highest values ever found in a galaxy centre
(e.g., Doyon et al. 1994).

The energy source of the huge IR luminosity is controversial.
Many IR spectroscopic studies
(e.g. Genzel et al. 1998; Ridgway et al. 1994; Rieke et al. 1985;
Weight et al. 1984)
have suggested that
main energy source of the IR emission is starburst activity,
which is presumably a super-starburst induced by a merger of
two galaxies (Joseph \& Wright 1985; Chevalier \& Clegg 1985).
The ground-based optical spectrum can be classified as LINER,
and is interpreted as a result of shock heating
(Heckman et al. 1987).
On the other hand, 
a significant contribution from an active galactic nucleus (AGN)
similar to Seyfert galaxies
was also discovered from IR spectroscopy (DePoy et al. 1986).
Another hint of an AGN in NGC~6240 is
the presence of compact bright radio cores
(Carral et al. 1990 but see Colbert et al. 1994).
{\sl HST} discovered a core that is excited higher than LINER
(Rafanelli et al. 1997).

X-ray observation provides an important tool for investigating both
the starburst and AGN activity.
The {\it ROSAT} (Tr\"{u}mper 1990) observations showed that NGC~6240 is 
fairly bright in the soft X-ray band below 2~keV
with a luminosity larger than $5\times10^{42}$ ergs s$^{-1}$
(0.1-2.0~keV), and
the bulk of the X-ray emission is extended in a scale of $\sim 25''$
(Schulz et al. 1998; Komossa, Schulz, \& Greiner 1998;
Iwasawa \& Comastri 1998).
Detailed spectroscopic studies with {\it ASCA} (Tanaka, Inoue, \& Holt 1994)
data of NGC~6240 have shown 
that the soft X-ray spectrum can be explained with two thermal components,
a cooler component with a temperature of 0.2--0.6~keV and 
a hotter component of $\sim1$~keV with an excess absorption
of $\sim10^{22}$ cm$^{-2}$ (Iwasawa \& Comastri 1998).
These results show that these soft X-rays are most likely originated 
 from thermal processes, which may arise from starburst activities
(e.g. Heckman et al. 1987).

On the other hand, in the 3--10~keV band, another very hard continuum
with a strong iron-K emission feature was observed with {\it ASCA}
(Mitsuda 1995; Iwasawa \& Comastri 1998).
The observed spectral feature can be accounted for in terms of
Compton reflection from optically thick material
(e.g. Lightman \& White 1988; George, Nandra \& Fabian 1990),
and is generally accepted as evidence for the presence of an AGN
in NGC~6240. The emission-line profile further indicates
that a part of the reflector is highly ionized
(Mitsuda 1995; Iwasawa \& Comastri 1998),
and both the {\it ROSAT} and {\it ASCA} spectra were modeled with 
a reflection component from warm material surrounding the AGN
(Netzer, Turner, \& George 1998; Komossa et al. 1998).

As described above, 
many observational facts of NGC~6240, in particular X-ray spectroscopy,
supports the presence of an AGN in NGC~6240,
which may account for a significant fraction of the huge IR luminosity.
However, the intrinsic power of the AGN is still uncertain,
The hidden AGN could be visible as a strongly absorbed X-ray continuum
above 10~keV, penetrating through a thick layer of the obscuring matter.

Very recently, Vignati et al. (1999) has published
the {\it BeppoSAX} results of NGC~6240,
concluding the detection of the direct X-rays from an AGN.
As described in this paper,
we performed an independent study of NGC~6240 with {\it RXTE}
which carries the PCA (Jahoda et al. 1996)
and HEXTE (Rothschild et al. 1998) covering 2--250~keV.

In the spectral analysis, we also utilized the {\it ASCA} data
covering 0.5--10~keV energy band
with two Gas Imaging Spectrometers (GIS: Ohashi et al. 1996)
and two X-ray CCD cameras (SIS: Burke et al. 1991).
Simultaneous use of the {\it ASCA} data provides more constraints in
modeling the spectrum.
Moreover, because {\it RXTE} has no imaging capability,
the {\it RXTE} spectrum may be subject to contamination by nearby sources.
In fact, we find from the {\it ASCA} image that it is the case, and
the {\it ASCA} data are used to correct the {\it RXTE} spectrum.

We have examined various spectral models, 
and conclude the presence of a high-luminosity AGN in NGC~6240.
Our results are in essential agreement with those of Vignati et al. (1999).

\section{{\it RXTE} data}
\subsection{Observation and data reduction}
The {\it RXTE} observation of NGC~6240 was performed
on 1997 Nov 9-11.
The data were reduced with Ftools 4.1 and 4.2,
We discarded the PCA and HEXTE data that were taken when
the centre of field of view was within 10$^{\circ}$
of the local horizon, and when it was off NGC~6240
by more than 0.01 degree.
We also selected the time period when all the five PCUs of PCA were on.
The total on-source exposure time after the data screening process
is 31.4 ksec and 10.3 ksec for the PCA and HEXTE, respectively.
All the five PCUs' top layers of PCA are used for the analysis,
while one of 8 detectors of HEXTE (\#3 in cluster B) is excluded
because of its failure.
We found no significant time variability from the PCA background-subtracted
light curve in the entire energy band (3-20~keV) 
as well as in several different energy bands.

\subsection{Derivation of PCA and HEXTE spectra}
We estimated the PCA background using the current standard method.
\footnote{The {\tt pcabackest} software was used
with the background model files of
{\tt pca\_bkgd\_faint240\_e03v02.mdl} and
{\tt pca\_bkgd\_faintl7\_e03v01.mdl}.}
We attempted to estimate systematic error associated with
the background subtraction by comparing the estimated background
with the on-source data in the range above 30 keV
where the source signal in PCA should be negligible due to low detection
efficiency.
The count rates of both agreed with each other within 0.01\%.
Therefore, we did not introduce any systematic error to the PCA 
background.

We estimated the HEXTE background using the data
taken during the off-target pointings (Rothschild et al. 1998).
In order to estimate the HEXTE-background systematics,
we accumulated spectra from the +1.5$^{\circ}$ off-pointings
and the --1.5$^{\circ}$ off-pointings, separately,
and found that the difference between the two background spectra
was within $\pm$2\%.
Therefore, we introduced a 2\% systematic error 
to the HEXTE background spectrum.

\begin{figure}
\centerline{
\psfig{file=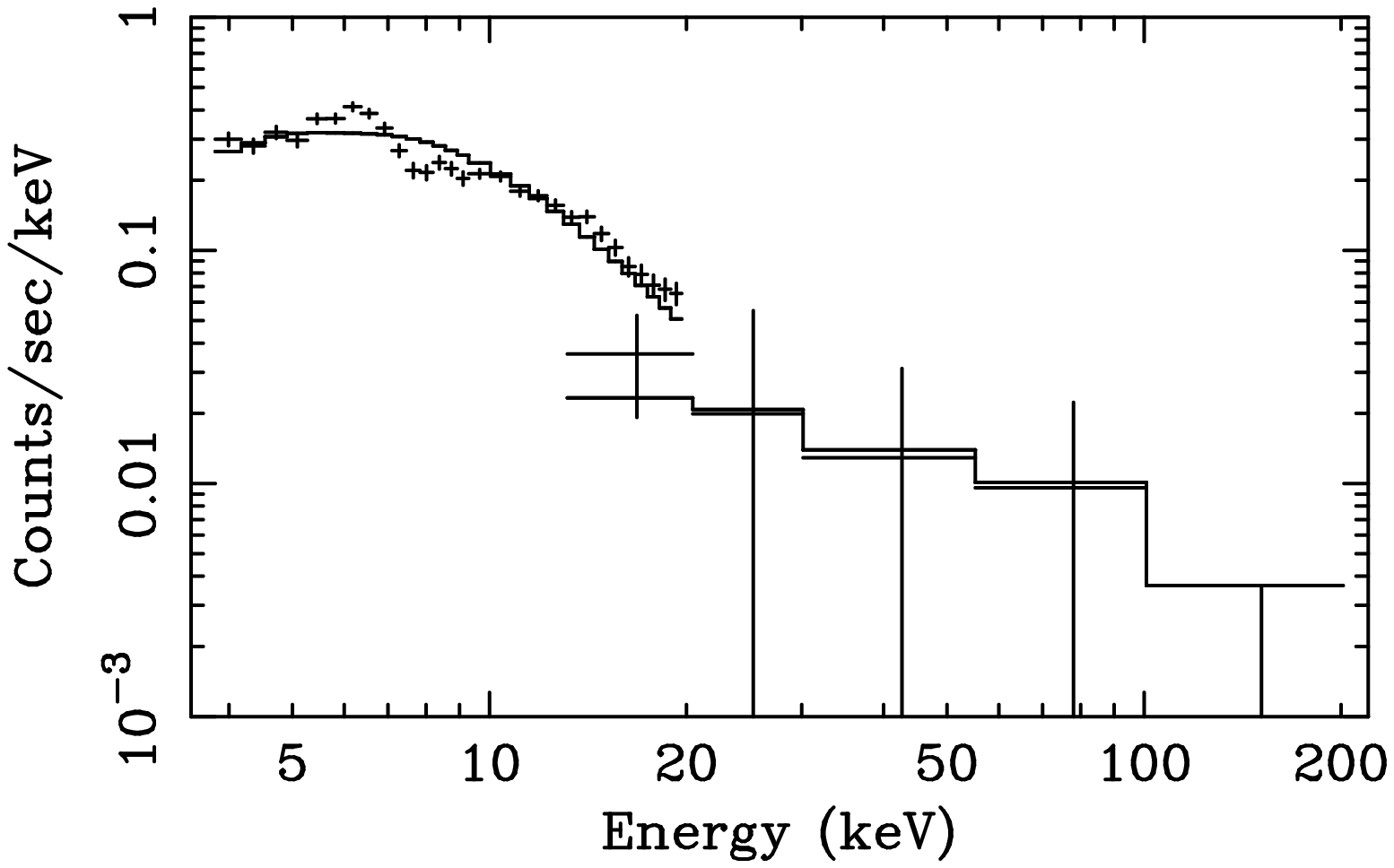,width=14.0cm,angle=-90,clip=}}
\vspace{-4.5cm}
\centerline{
\psfig{file=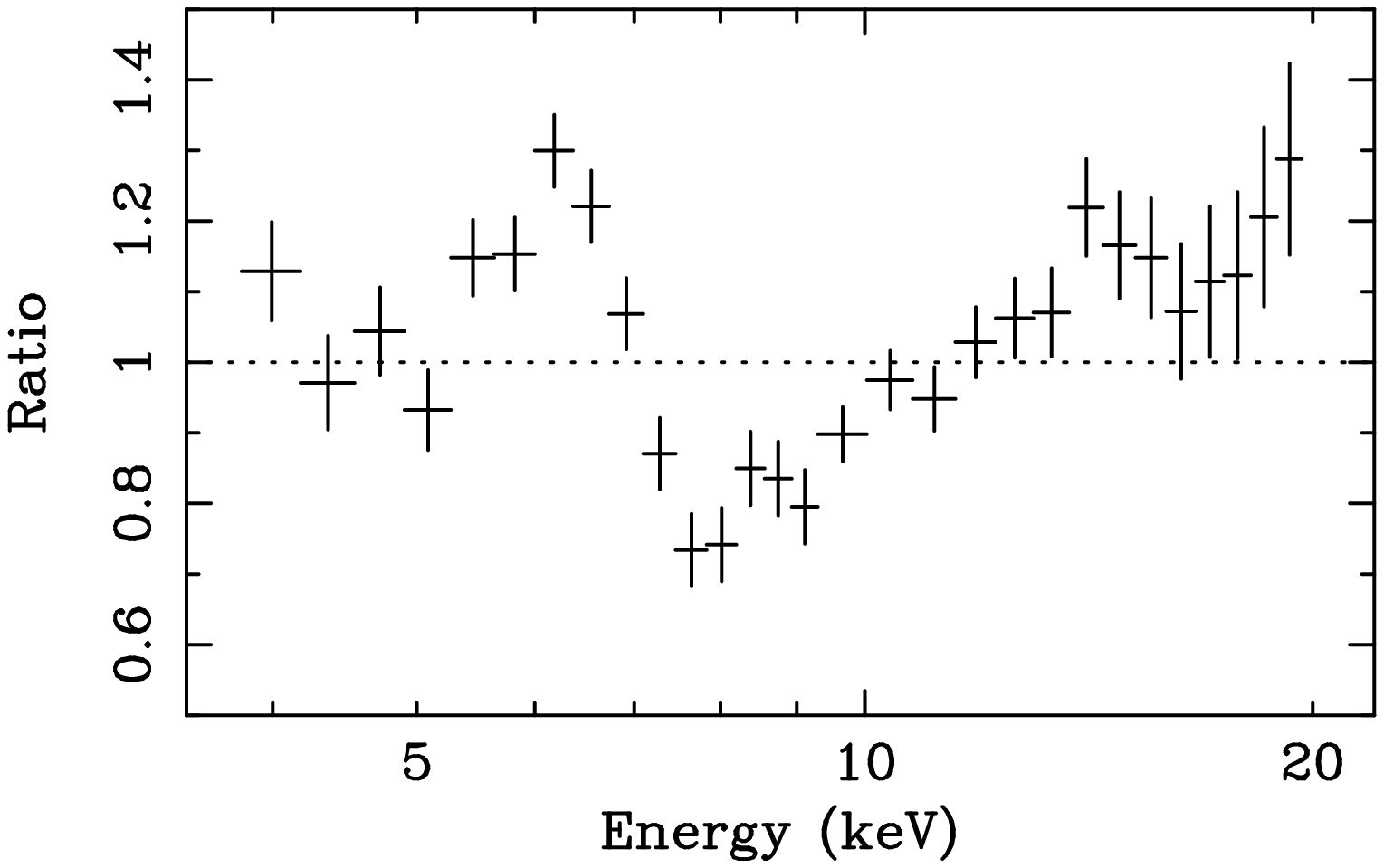,width=14.0cm,angle=-90,clip=}}
\vspace{-4cm}
\caption{(a) The background-subtracted spectrum obtained 
with PCA and HEXTE (crosses), fitted with a single power-law model
convolved with the energy-response matrices (solid lines).
(b) The ratio between the PCA data and the best-fit power-law model.}
\label{fig:rxtespec}
\end{figure}

Figure~\ref{fig:rxtespec}a
illustrates the PCA and HEXTE energy spectra thus obtained.
Significant source signals are detected up to $\sim$25keV
in the PCA data. However, the PCA data below 4~keV and above 20~keV 
were discarded for the spectral analysis below, 
due to the uncertainty of the energy response reported
by Gierli\'{n}ski et al. (1999) and Jahoda et al. (1996).
The HEXTE spectrum shows significant signals up to 20~keV,
while only upper limits are obtained above 20~keV.

\subsection{Spectral analysis}
In order to see the overall spectral characteristics,
we first fit the spectra with a simple power-law model modified
by Galactic absorption, i.e.
$f(E) = K E^{-\Gamma} e ^{ -\sigma_{\rm ph} N_{\rm H,Gal} } $,
where $E$ is X-ray energy, f(E) is photon flux given in units 
of photons s$^{-1}$ cm$^{-2}$ keV$^{-1}$,
$\Gamma$ is photon index, $K$ is the normalization factor,
$\sigma_{\rm ph}$ is the cross-section of photoelectric
absorption given by Morrison \& McCammon (1983),
and $N_{\rm H,Gal}$ is the Galactic hydrogen column density
set equal to $5.8\times10^{20}$ cm$^{-2}$ (Dickey \& Lockman 1990).
Using XSPEC software (ver. 10.0), we performed a minimum-chi-square
fitting.
The best-fit photon-index is 0.51 yielding
$\chi^2/\nu$ = 210/32 (Figure~\ref{fig:rxtespec}a).
The ratio of the data to the best-fit power-law model
(Figure~\ref{fig:rxtespec}b)
clearly shows evidence for an iron K emission line at $\sim$6~keV
and a flatter continuum above $\sim$8~keV.

\begin{figure}
\centerline{
\psfig{file=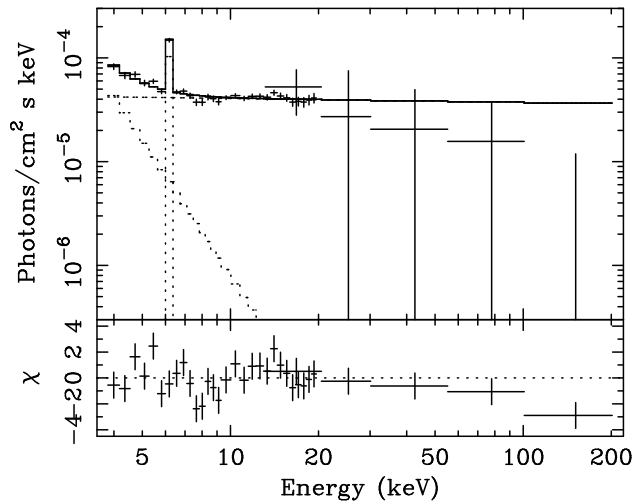,width=14.0cm,angle=-90,clip=}}
\vspace{-2cm}
\caption{The unfolded energy spectrum
(detector response was approximately removed)
of the PCA and HEXTE data (crosses)
fitted with two power-law model and an emission line.
Each model component is shown with dotted lines and 
the total model is the solid line.}
\label{fig:rxte2powfit}
\end{figure}

Then, we modeled the spectrum with two power-law continua
(soft and hard) plus a line emission, as expressed by
$ f(E) = ( K_s E^{-\Gamma_s} + K_h E^{-\Gamma_h} + line (K_l, E_l) )
e ^{ -\sigma_{\rm ph} N_{\rm H,Gal} } $.
The iron-line central energy ($E_l$) and the intensity ($K_l$) are
free parameters, and the line width is assumed to be zero.
The fit result is shown in Figure~\ref{fig:rxte2powfit}
and the best-fit parameters are summarized in Table~1.
Notable properties of the best-fit model are 
very small photon index, $\sim$0, of the hard pawer-law continuum, 
and a large equivalent width, 0.79~keV, of the iron line.
An absorption edge structure is noticeable at 7--8~keV
in the fit residual (Fig.~\ref{fig:rxte2powfit}), 
which is most likely the iron K-edge.
These features are characteristic of the reflected X-rays from an 
optically thick material as has been pointed out by several authors
(Iwasawa \& Comastri 1998; Netzer et al. 1998).
X-rays impinging on optically thick matter are photo-absorbed 
as well as Compton-scattered.
These processes form a very flat continuum around $\sim10$~keV, 
together with the K-absorption edge and K emission line of iron.
We fit the {\it RXTE} spectrum together with {\it ASCA} data in \S~4 with
models including the Compton reflection.

\begin{table}
\label{tab:1}
\begin{center}
  \caption{The {\it RXTE} spectral fit}
  \begin{tabular}{ll}
    \hline 
    \hline 
    Parameter                        & Value   \\
    \hline 
    $K_{\rm s}$(counts s$^{-1}$ cm$^{-2}$ at 1 keV) &  $1.83\times10^{-2}$\\
    $\Gamma_{\rm s}$                       &  4.4$^{+1.4}_{-1.2}$    \\
    $K_{\rm h}$(counts s$^{-1}$ cm$^{-2}$ at 1 keV) &  $4.45\times10^{-5}$\\
    $\Gamma_{\rm h}$                       &  0.04$^{+0.09}_{-0.10}$ \\
    $K_l$(counts s$^{-1}$ cm$^{-2}$)$^{a)}$ &  $3.86\times10^{-5}$ \\
    $E_l$(keV)                             &  6.34 $\pm0.09$ \\
    $\chi^2/\nu$                           &  48.0/28 \\
    \hline 
  \end{tabular}
\end{center}
\begin{itemize}
\setlength{\itemsep}{2mm}
  \item[$^{a)}$]: The physical line width is assumed to be zero.
	The equivalent width is calculated to be 0.79~keV.
\end{itemize}
\end{table}

\section{ASCA data}
\subsection{Observation and data reduction}
{\it ASCA} observation of NGC~6240 was performed on 27 March, 1994.
The GIS provides an X-ray image with a circular field of view of
$\sim50'$ diameter.
The SIS was operated with 2-CCD mode, covering
a much smaller, rectangular field of view of $\sim 11' \times 22'$.
Iwasawa \& Comastri (1998) presented the result of the SIS data only. 
Here, we analyse both the SIS and GIS data.
In particular, the GIS data are useful in order to look for any 
contamination sources in the field around NGC~6240.

We accepted the GIS and SIS data that were taken when 
the X-Ray Telescope (XRT: Serlemitsos et al. 1995) axis
was more than 5$^{\circ}$ above the local horizon,
and when the geomagnetic cutoff rigidity was larger than 6 GeV/c
in order to ensure a low and stable background.
Additional screening condition that the elevation angle
 from the sunlit earth is greater than 25$^{\circ}$ and 20$^{\circ}$
was applied to the SIS0 and SIS1 data, respectively.
As reported by Turner et al. (1997),
the {\it ASCA} data of NGC~6240 show no significant time variation.

\subsection{GIS image}
The GIS image in the full energy band 0.5--10~keV shows a
point-like source that coincides with NGC~6240 
within $\sim1$ arcmin systematic error of the satellite attitude.
In addition to the emission from NGC~6240,
the soft Galactic diffuse emission covers 
the entire GIS field of view.
The position of NGC~6240, $(l,b)$=$(20.73, 27.29)$,
is in the middle of an excess-emission structure of Loop I, which is 
found in the {\it ROSAT} All Sky Survey (Snowden et al. 1997).
The energy spectrum of the excess emission
is very soft ($kT\sim0.3$ keV), and it is significant only
in the soft energy band below $\sim2$~keV.

\begin{figure}
\centerline{
\psfig{file=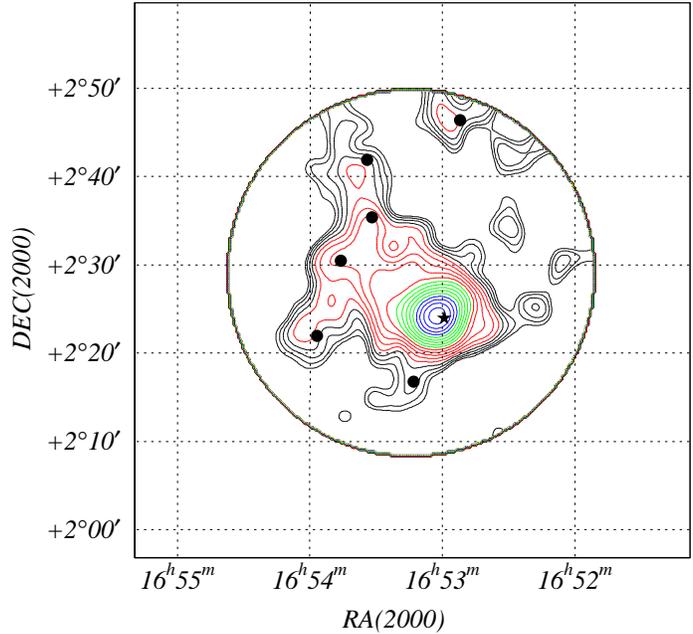,width=12.0cm,angle=0,clip=}}
\caption{The 2.4--10~keV X-ray intensity contour derived
with the GIS data is shown in logarithmic scale.
After the background (NXB + CXB) is subtracted,
the image was smoothed with a Gaussian of $\sigma=1.25'$.
The outer most solid circle shows the edge of the GIS field of view.
The position of NGC~6240 is marked with $\star$, while
$\bullet$ show positions of point like sources detected 
with {\it ROSAT} PSPC.}
\label{fig:gisimage2}
\end{figure}

On the other hand, at higher energies,
we find an extended structure as clearly seen in the image for
the 2.4--10~keV band shown in Figure~\ref{fig:gisimage2}, in which
the background consisting of non-X-ray background (NXB) and
cosmic X-ray background (CXB) has been subtracted.
These are estimated from the data obtained in
the blank sky observations and data taken when the XRT
is pointing at the dark (night) earth
(for detail see Ikebe et al. 1995; Ishisaki et al. 1997).
The background was subtracted to derive the image.
In Figure~\ref{fig:gisimage2},
notable emission is located north-east of NGC~6240,
and extends towards north as well as east.
Since the extension of the emission is much larger than that of
the soft X-ray emission of NGC~6240 detected with {\it ROSAT},
we consider it to be an unassociated with NGC~6240.
In the {\it ROSAT} PSPC data, there exist several faint sources
around NGC~6240 as illustrated in Figure~3.
These sources, if blurred with the angular response of the GIS,
seem to form a brightness structure consistent with that observed
with the GIS.
Therefore, the extended X-ray structure is most probably due to
these faint background sources.
Since the field of view of {\it RXTE}/PCA and HEXTE are
both $\sim$1 degree FWHM,
the emission from the contamination sources will contribute
to the {\it RXTE} spectra significantly.
Below we examine the {\it ASCA} spectrum of NGC~6240 and the properties of 
these contamination sources.

\subsection{Spectrum of NGC~6240}
We construct the GIS energy spectrum of NGC~6240 from a circular
region of 3 arcmin radius centered on the X-ray peak.
The spectra from the two GIS sensors, GIS-2 and GIS-3, are summed.
The NXB + CXB background is subtracted, as has been done for the GIS image.
The soft excess emission is not subtracted, but it is negligible 
above 2.0~keV.

The SIS spectrum of NGC~6240 is extracted within a circle
of 3 arcmin radius from the SIS-0 chip-1 and the SIS-1 chip-3.
The data taken from the two chips are summed together.
For the SIS background, we accumulate photons
 from a region where no contamination source is present  
on the same CCD chips with which NGC~6240 was observed.
Therefore, the soft Galactic diffuse component is also subtracted
as a part of the background.

\begin{figure}
\centerline{
\psfig{file=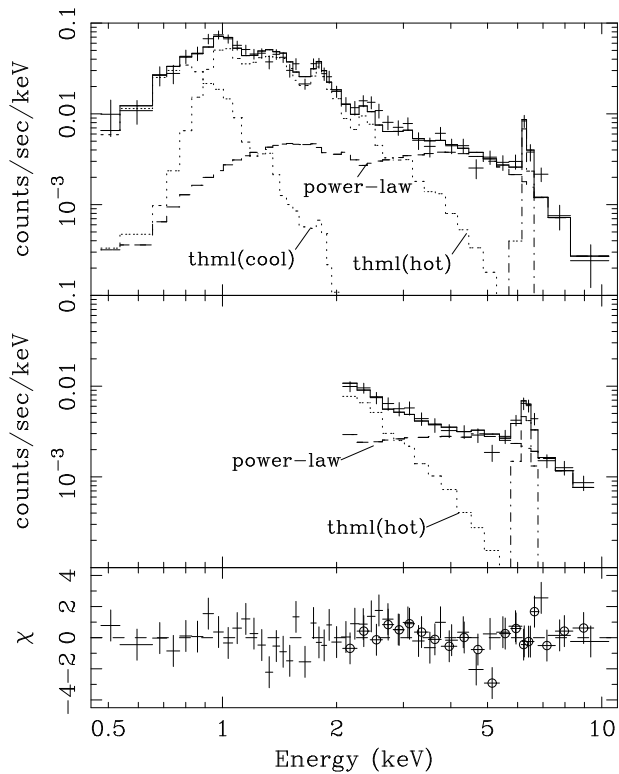,width=14.0cm,angle=-90,clip=}}
\caption{(top) The background-subtracted SIS spectra (crosses)
fitted with sum of thermal component (dotted lines), 
an emission line (dot-dashed line), and a power-law model (dashed line).
All the model components are convolved with the energy response
matrices.
The solid line shows the sum of all the model components.
(middle) The same as the top panel, but for the GIS data.
(bottom) The fit residuals for the SIS data (crosses) 
and the GIS data (crosses with an open circle).}
\label{fig:ascaspec}
\end{figure}

The GIS and SIS spectra thus obtained are illustrated in 
Figure~\ref{fig:ascaspec}.
Following Iwasawa \& Comastri (1998),
we fit the {\it ASCA} spectrum of NGC~6240 with a two-component model
that includes a thermal component and an AGN component.
The thermal component consists of emission from two 
optically-thin thermal plasmas of different temperatures,
where the higher temperature component has an excess absorption.
The AGN component consists of an absorbed power-law continuum
and line emission.
The model can be written as;
\begin{eqnarray}
f(E) & = & \{ thml(T_c,Z,K_c)
	+ e^{-\sigma_{\rm ph} N_{\rm H}}\ thml(T_h,Z,K_h) \nonumber\\
 & + & line(E_l,K_l) 
+ e^{-\sigma_{\rm ph} N_{\rm H,AGN}}\ K_{\rm AGN} E^{-\Gamma} \} \nonumber\\
 & \times & e^{-\sigma_{\rm ph} N_{\rm H,Gal}} \ ,
\label{eq:powlaw}
\end{eqnarray}
where the parameters in parentheses are kept free.
For the thermal plasma emission code, we employ the Mewe-Kaastra model
(Mewe, Gronenschild, \& van den Oord 1985; 
Mewe, Lemen, \& van den Oord 1986; Kaastra 1992)
modified by Liedahl, Osterheld, \& Goldstein (1995),
which is implemented in XSPEC as the MEKAL model.
The metallicities, $Z$, of the two thermal components
are assumed to be the same.
The abundance ratios among different elements are fixed to
be the fiducial solar values given by Anders \& Grevesse (1989).

With this model, we fit the GIS and SIS spectra simultaneously.
In this simultaneous fit, the GIS data below 2.0~keV are excluded because
of a slight uncertainty in the energy scale of the GIS arising from
the complex xenon M-edge structure in this range.
A good fit is obtained as shown in Figure~\ref{fig:ascaspec} 
and Table~2,
and the results are essentially the same 
as obtained by Iwasawa \& Comastri (1998) using the SIS data only.
As noted by Iwasawa \& Comastri (1998),
the striking spectral features are a very flat continuum
above $\sim$4~keV with a photon index of $\sim 0$
and a strong iron K-line with an equivalent width of 1.2~keV.

\begin{table}
\label{tab:2}
\begin{center}
  \caption{ASCA spectral fit$^{a)}$}
  \begin{tabular}{ll}
    \hline 
    \hline 
    Parameter                        & Value \\
    \hline 
    $K_c$ (cm$^{-3}$)$^{b)}$         & $3.8\times10^{64}$   \\
    $T_c$ (keV)                      & 0.31$^{+0.11}_{-0.05}$ \\
    $K_h$ (cm$^{-3}$)$^{b)}$         & $3.5\times10^{65}$    \\
    $T_h$ (keV)                      & 1.07$^{+0.24}_{-0.10}$  \\
    $Z$   (solar)                    & 1.41$^{+11.18}_{-0.84}$ \\
    $N_{\rm H}$ ($10^{22}$cm$^{-2}$) & 0.92$^{+0.14}_{-0.16}$ \\
    $K_l$(counts s$^{-1}$ cm$^{-2}$)$^{c)}$ & $2.9\times10^{-5}$  \\
    $E_l$(keV)                       & 6.46$^{+0.03}_{-0.04}$ \\
    $N_{\rm H,AGN}$ ($10^{22}$cm$^{-2}$) & 0.0$^{+2.4}_{-0}$ \\
    $K_{\rm AGN}$ (counts s$^{-1}$ cm$^{-2}$ keV$^{-1}$ @1keV)  
				     & $2.0\times10^{-5}$   \\
    $\Gamma$                         & -0.10$^{+0.29}_{-0.43}$ \\
    $\chi^2/\nu$                     & 59.4/55 \\
    \hline 
  \end{tabular}
\end{center}
\begin{itemize}
\setlength{\itemsep}{2mm}
  \item[$^{a)}$]: Throughout this paper, 
	for fitting the SIS spectrum only,
	an additional absorption with equivalent hydrogen column density
	of $3.0\times10^{20}$ cm$^{-2}$ was artificially applied,
	in order to account for a systematic overestimate of
	the absorption column density
	(Dotani et al. 1996; Cappi et al. 1998).
  \item[$^{b)}$]: Emission integral.
  \item[$^{c)}$]: The physical line width is assumed to be zero.
	The equivalent width with respect to the power law continuum
	is 1.24 keV.
\end{itemize}
\end{table}

\subsection{Contamination sources' spectrum}
\begin{figure}
\centerline{
\psfig{file=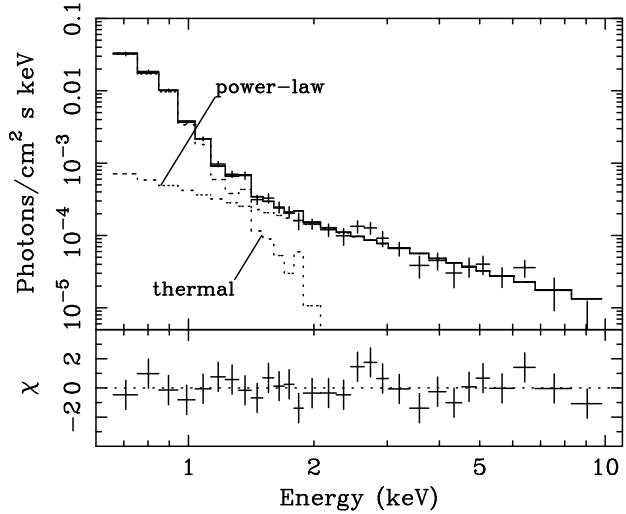,width=14.0cm,angle=-90,clip=}}
\caption{(top) The unfolded energy spectrum of the contamination sources obtained
with GIS.
The spectrum is fitted with sum of a thermal and a power-law component
(dotted lines).
Solid line shows the total model.
(bottom) The fit residuals.}
\label{fig:contamispec}
\end{figure}

Although the contamination sources near NGC~6240 are not identified,
we only need the energy spectrum for the purpose of the present work.
We construct the spectrum from the whole field of view of the GIS
excluding the 3-arcmin radius circle centered on the X-ray peak of
NGC~6240, from which the estimated background (NXB + CXB) is subtracted.
In addition, because of an extended outskirts of the XRT point-spread function,
the spectrum still contains a significant contribution from NGC~6240.
Using a ray-tracing simulation software,
the contribution from NGC~6240 outside the 3-arcmin radius 
circle is estimated and subtracted.
Figure~\ref{fig:contamispec} shows the GIS spectrum thus obtained.
The resultant GIS spectrum contains
the total photons from the contamination sources within
the GIS field of view and the Galactic soft diffuse emission. 

As shown in Figure~\ref{fig:contamispec},
thus obtained GIS spectrum is fitted satisfactorily ($\chi^2/\nu=18.4/25$)
with the sum of a thin-thermal model (MEKAL model) and a power-law model. 
For the thermal component,
we obtain the best-fit temperature of 0.23$\pm0.04$~keV
and the surface brightness of 1.0$\times10^{-14}$ ergs s$^{-1}$ cm$^{-2}$
arcmin$^{-2}$ (0.5--2~keV),
assuming the element abundances to be solar. 
We identify this thermal component to be the Galactic diffuse emission,  
since these values are consistent with the {\it ROSAT} results
(Snowden et al. 1997).
Thus, the spectrum of the contamination sources can be expressed by a
power law with the best-fit photon index of 1.56$\pm$0.23.
Its flux is
2.1$\times10^{-12}$ ergs s$^{-1}$ cm$^{-2}$ (2 -- 10~keV),
which is $\sim90$\% as high as that of NGC~6240.
Therefore, the contribution from the contamination sources
to the {\it RXTE} spectrum is quite substantial, and
will be taken into account for the analysis of the {\it RXTE} spectrum
in the next section.
It is to be noted that the spectrum of the contamination sources
(added together) is pretty hard, consistent with that of AGN.
This suggest that these sources, if not all, are possibly AGN.

\section{RXTE + ASCA joint fit}
In this section, we model the 0.5--200~keV energy spectrum of NGC~6240
obtained with {\it RXTE} and {\it ASCA},
taking into account the contamination sources detected in the GIS
field of view.
Below 3~keV, the X-ray emission is dominated by the thermal
component that presumably originates from the starburst activity
in the galaxy, 
and is well described with the two temperature model 
(Iwasawa \& Comastri 1998) as shown in \S~3.3.
The X-ray emission from the AGN in NGC~6240 dominates
above 4~keV and we perform detailed modeling below.

\subsection{Power law model}
For modeling the energy spectrum of the AGN,
we begin with the simplest model, power-law + line,
given by equation~\ref{eq:powlaw} in \S~3.3,
The {\it ASCA} GIS and SIS spectra of NGC~6240, the PCA spectrum,
and the HEXTE spectrum were fitted simultaneously.
Since the X-ray intensity of the AGN may have varied
between the {\it ASCA} and {\it RXTE} observations performed at different periods,
the normalization factor of the power-law component, $K_{\rm AGN}$,
is left free for the {\it ASCA} and {\it RXTE} spectra, respectively.
Since the iron K-line is presumably generated
by reprocessing of the X-rays from the AGN on an optically thick matter,
the line intensity will be proportional to the AGN luminosity
on the time average.
Therefore, we assumed that
the ratio of the flux of the iron line to that of
the power-law continuum, $K_l/K_{\rm AGN}$,
was the same in the two separate observations 
with {\it ASCA} and {\it RXTE}, respectively.
On the other hand, the normalization factor of the thermal
components, $K_c$ and $K_h$, are assumed to be common to
the {\it RXTE} and {\it ASCA} spectra, since the thermal component
which is extended should not vary with time.
All other parameters are also tied between the two observations.
In order to take into account the hard contribution source,
a power-law model with the photon index 1.56 as derived in the previous 
section is included in the PCA and HEXTE spectra with a free
normalization factor.
We assume that the power law spectrum extends beyond 10~keV with the 
same photon index.

\begin{figure}
\centerline{
\psfig{file=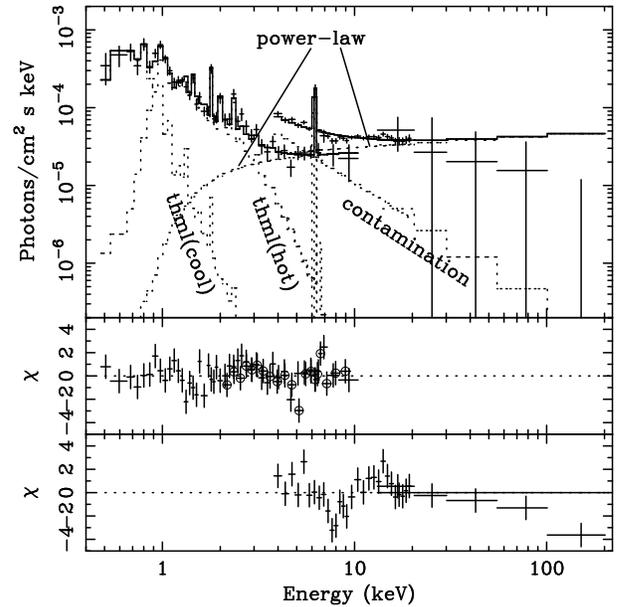,width=14.0cm,angle=-90,clip=}}
\caption{The result of the simultaneous fit to the {\it ASCA}/GIS,
{\it ASCA}/SIS,
{\it RXTE}/PCA, and {\it RXTE}/HEXTE with sum of thermal component, emission line,
and a power-law model.
(top) The unfolded spectral data (crosses) of {\it ASCA}/SIS, {\it RXTE}/PCA and
{\it RXTE}/HEXTE.
Individual fitting model components are shown in dotted lines,
while the total is given with the solid line.
The GIS data are omitted for the display.
(middle) The fit residuals for the SIS data (crosses) and 
for the GIS data (crosses with open circle).
(bottom) The fit residuals for the PCA and HEXTE data.
}
\label{fig:powfit}
\end{figure}

As shown in Figure~\ref{fig:powfit} and Table~\ref{tab:3},
the fit is not acceptable.
As expected, the obtained parameters of the thermal component are
essentially the same as those given in Section 3.3, since 
the {\it RXTE} spectrum has little influence below 4~keV.
However, the best-fit power-law continuum of AGN is significantly flatter
than that given by Iwasawa \& Comastri (1998).
Furthermore, the residuals require an even harder continuum
in 8--15~keV band,
and also suggests a spectral steepening above $\sim30$~keV.
These features together with a large equivalent width of Fe K-line 
are consistent with a Compton reflection spectrum. 
We therefore incorporate Compton reflection in the next subsection.

\begin{table}
\label{tab:3}
\begin{center}
  \caption{Power law model}
  \begin{tabular}{ll}
    \hline 
    \hline 
    Parameter                        & Value               \\
    \hline 
    $K_l(ASCA)$(counts s$^{-1}$ cm$^{-2}$)$^{a)}$ & $2.8\times10^{-5}$  \\
    $E_l$(keV)                       & 6.44                \\
    $N_{\rm H,AGN}$ ($10^{22}$cm$^{-2}$) & 1.1             \\
    $\Gamma$                         & -0.15               \\
    $K_{\rm AGN}$(ASCA)$^{b)}$       & $1.9\times10^{-5}$  \\
    $K_{\rm AGN}(RXTE)/K_{\rm AGN}(ASCA)$ 
				     & 1.16                \\
    $\chi^2/\nu$                     & 131.5/87 \\
    \hline 
  \end{tabular}
\end{center}
\begin{itemize}
\setlength{\itemsep}{2mm}
  \item[$^{a)}$]: The physical line width is assumed to be zero.
	The equivalent width with respect to the power-law continuum
	is 1.13~keV.
  \item[$^{b)}$]: In the unit of counts s$^{-1}$ cm$^{-2}$ keV$^{-1}$ @1keV.
\end{itemize}
\end{table}

\subsection{Reflection model}
We employ the Compton reflection model developed
by Magdziarz \& Zdziarski (1995) implemented in XSPEC
as the PEXRAV model.
Using the PEXRAV model, we can calculate the reflected X-ray spectrum
when X-rays of a power-law spectrum illuminate an optically thick 
layer of material that is predominately 
neutral except hydrogen and helium.
Here, we assume that the direct X-rays from the AGN are totally blocked
by a thick absorber on the line of sight.
The parameters that describe the model are:
$\Gamma$, photon index of the incident power law spectrum with an 
exponential cutoff at $E_c$;
$R=\Omega/2\pi$, where $\Omega$ is the solid angle subtended by
the optically thick material;
$\mu=cos\theta$, where $\theta$ is the angle between the line of sight and 
the normal vector of the optically thick layer; 
$A_{\rm Fe}$, the iron abundance of the optically thick material.
Element abundances besides iron are assumed to be the solar values.

In the following fitting, $E_c$ is fixed at 200~keV (practically a
single power law in the observed range).
$\mu$ is fixed at 0.45 so that the model spectrum is closest 
to that averaged over all viewing angles as described in
Magdziarz \& Zdziarski (1995).
Combining with the thermal component and the line emission,
the overall fitting model is given by
\begin{eqnarray}
f(E) & = & \{ thml(T_c,Z,K_c)
	+ e^{-\sigma_{\rm ph} N_{\rm H}}\ thml(T_h,Z,K_h) \nonumber\\
    & + & line(E_l,K_l) + refl(\Gamma,K_{\rm AGN}) \}
	e^{-\sigma_{\rm ph} N_{\rm H,Gal}} \ ,
\label{eq:ref}
\end{eqnarray}
where free parameters are shown in parentheses.
The normalization factors of the reflection component, $K_{\rm AGN}$,
for the {\it ASCA} and {\it RXTE} spectra are left free from each other 
for possible time variation of the AGN luminosity.
The flux ratio between the Fe K-line and the AGN continuum 
$K_l/K_{\rm AGN}$ is assumed to be constant.
All other parameters are also assumed to be the same
between the two observations.
Another power-law with a photon index of 1.56 that represents the
contamination sources is added to the model for the PCA and HEXTE spectrum.

\begin{figure}
\centerline{
\psfig{file=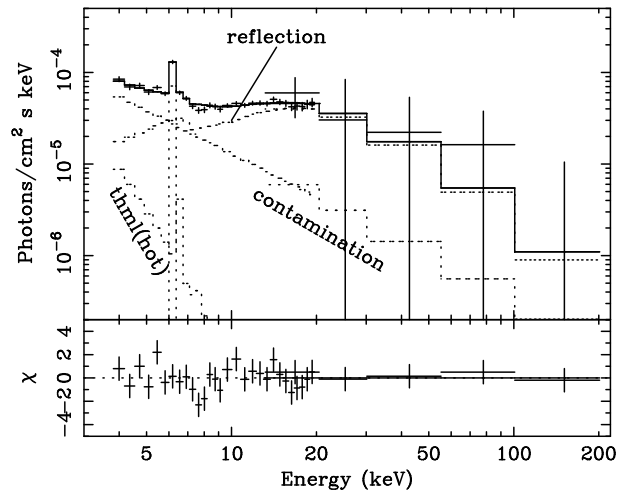,width=14.0cm,angle=-90,clip=}}
\caption{Unfolded spectrum of the {\it RXTE} data (crosses)
fitted with sum of the thermal component, reflection component,
and a power-law for the contamination source.
Each model component is shown with dotted lines, while the total
is illustrated with the solid line.
Although the {\it ASCA}/GIS and SIS data are not shown here,
they are also fitted jointly.}
\label{fig:reffit}
\end{figure}

As shown in Figure~\ref{fig:reffit} and Table~4,
this model can account for the 0.5--200~keV wide-band spectrum
satisfactorily.
The flux of the contamination sources in the {\it RXTE} spectrum determined 
 from the fit is in good agreement with that obtained with the GIS
within $\sim10\%$.
The two normalization factors of the AGN component
obtained from the {\it ASCA} and {\it RXTE} spectra respectively
happened to be very close ($K^{\rm RXTE}/K^{\rm ASCA}$=0.92--1.26),
as were the Fe K-line intensities nearly equal in these two
observations.
The equivalent width of the iron line with respect to the reflection
continuum is 0.8~keV,
which is a typical value for the fluorescence line associated with 
the Compton-reflection.

However, the best-fit photon index $\Gamma$ of the incident AGN 
spectrum is 1.26$\pm$0.13,
which is unusually small compared to the typical values 
for AGNs (e.g. Mushotzky, Done, \& Pounds 1993).
In this fit, the viewing angle $\mu = cos \theta$ was fixed at 0.45.
Even if we allow $\mu$ to vary (Table~4), 
photon index larger than 1.54 is not acceptable.
The intrinsic luminosity of the AGN is estimated to be larger
than $4.3\times10^{43}$ ergs/s, the value corresponding to the
maximum solid angle, i.e. $R=\Omega/2\pi$, of 1.0.

\begin{table}
\label{tab:4}
\begin{center}
  \caption{reflection model}
  \begin{tabular}{lll}
    \hline 
    \hline 
    Parameter               & Value (fit 1)$^{a)}$   & Value (fit 2)$^{a)}$ \\
    \hline 
    $\Gamma$                & 1.26$^{+0.13}_{-0.13}$ & 0.98$^{+0.56}_{-0.22}$\\
    $\mu = cos \theta$      & 0.45 (fix)    & $^{b)}$0.08$^{+0.87}_{-0.03}$\\
    Fe abundance (solar)    & 0.47$^{+0.22}_{-0.19}$ & 0.72$^{+0.38}_{-0.32}$\\
    $K_{\rm AGN}$(ASCA)$^{c)}$ & 2.7$\times10^{-3}$  & 4.6$\times10^{-3}$ \\
    $K_l(ASCA)$(counts s$^{-1}$ cm$^{-2}$)$^{d)}$ 
                             & 2.5$\times10^{-5}$     & 2.4$\times10^{-5}$ \\
    $E_l(ASCA)$(keV)         & 6.44$^{+0.04}_{-0.05}$ & 6.44$\pm0.04$ \\
    $K_{\rm AGN}(RXTE)$/$K_{\rm AGN}(ASCA)$         
                           & 1.08$^{+0.18}_{-0.16}$ & 1.05$^{+0.16}_{-0.15}$ \\
    $\chi^2/\nu$             & 97.2/87                & 96.4/86   \\
    \hline 
  \end{tabular}
\end{center}
\begin{itemize}
  \item[$^{a)}$]: $\mu$ is fixed or left free in the fit 1 and 2, respectively.
  \item[$^{b)}$]: The lower and upper limits correspond to
              the limitation of the model code.
  \item[$^{c)}$]: In the unit of counts s$^{-1}$ cm$^{-2}$ keV$^{-1}$ @1keV.
  \item[$^{d)}$]: The physical line width is assumed to be zero.
	Equivalent width with respect to the reflection continuum is
        0.88~keV and 0.84~keV for the fit 1 and 2, respectively.
\end{itemize}
\end{table}

\subsection{Reflection + Transmission model}
Although, the above reflection-only model gives a satisfactory fit
to the {\it RXTE} + {\it ASCA} data, the small photon index derived in \S~4.2
still remains to be problem.
As shown below, this is resolved by adding an absorbed AGN continuum
that is transmitted through a thick absorber on the line of sight.
(Here we do not consider the extent of the absorber,
hence we assume no scattering of X-rays into the line of sight.)
The transmitted AGN component is given by
\begin{equation}
e^{-N_{\rm H,AGN}(\sigma_{\rm ph}+\sigma_{\rm Th})}
K_{\rm AGN} E^{-\Gamma} e^{-E/E_{\rm c}}\ ,
\label{eq:tra}
\end{equation}
where $\sigma_{\rm Th}$ is the Thomson scattering cross-section,
and $N_{\rm H,AGN}$ is the hydrogen-column density along line of sight.
The attenuation by the Thomson scattering becomes important 
above $\sim10$~keV as compared to photoelectric absorption.

Adding the transmitted AGN component given 
by eq.~\ref{eq:tra} to eq.~\ref{eq:ref},
we construct
the {\it reflected- and transmitted-}AGN + thermal model as;
\begin{eqnarray}
f(E) & = & \{ thml(T_c,Z,K_c) + 
	e^{-\sigma_{\rm ph} N_{\rm H}}\ thml(T_h,Z,K_h)     \nonumber\\
  & + & line(K_l,E_l) + refl(\Gamma,R,E_c,K_{\rm AGN})   \nonumber\\
  & + & e^{-N_{\rm H,AGN}(\sigma_{\rm ph}+\sigma_{\rm Th})}
	K_{\rm AGN} E^{-\Gamma} e^{-E/E_{\rm c}} \} \nonumber\\
  & \times & e^{-\sigma_{\rm ph} N_{\rm H,Gal}}\ .
\label{eq:reftra}
\end{eqnarray}
The photon-index $\Gamma$, the cutoff energy, $E_C$ (fixed to 200~keV),
and the normalization, $K_{\rm AGN}$, are the same for 
both the reflection and transmitted components.
The relative intensities between the transmitted component 
and the reflection component is determined by
the solid angle, $R$, which is left free.
The metallicity of the elements lighter than Fe 
and the viewing angle $ \mu = cos \theta $ 
are fixed at 1.0 solar and 0.45, respectively.
As in the previous fits,
including the contamination-source component in the {\it RXTE} model,
we fit the four spectra jointly with this model.

\begin{figure}
\centerline{
\psfig{file=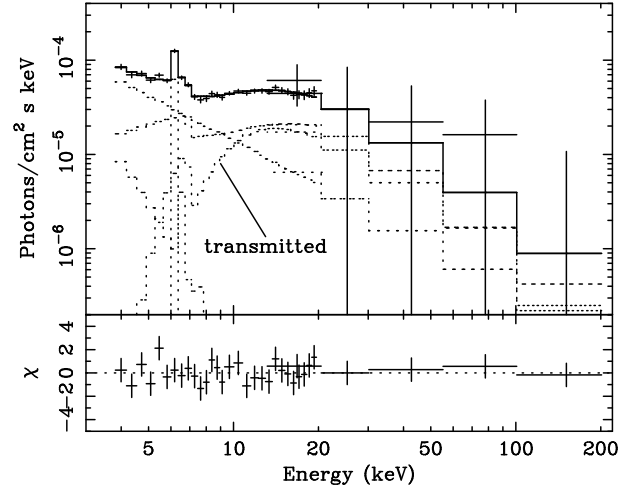,width=14.0cm,angle=-90,clip=}}
\caption{
The fit result of the reflection-and transmitted-AGN model.
Crosses are the unfolded spectral data.
Each model component is shown with dotted lines,
while the total is given with solid line.
Although the {\it ASCA}/GIS and SIS data were also
fitted jointly, only the {\it RXTE} data are shown here.}
\label{fig:reftra}
\end{figure}

The {\it reflected- and transmitted-}AGN + thermal model
gives a good fit (Figure~\ref{fig:reftra}, Table~5).
The derived photon index is $\Gamma$= 1.59 $^{+0.42}_{-0.26}$,
which is consistent with the canonical value of 1.9 (Pounds et al. 1990).
The best-fit solid angle of the reflector, $R = \Omega/2\pi$, is 0.51,
and the absorption column density, $N_{\rm H,AGN}$, is found
to be $1.7\times10^{24}$ cm$^{-2}$.
Since $R$ was poorly constrained,
the intrinsic luminosity of the AGN has a large error range.
If we assume that $R$ does not exceed 1.0,
$L_{\rm X}$ is obtained to be $1.1^{+4.5}_{-0.5}\times10^{44}$ ergs/s in
the range 2 -- 10~keV.
When we leave the viewing angle $\mu$ free,
the fitting parameters as well as the intrinsic luminosity
remain essentially the same, if $R \leq 1.0$ (Table~\ref{tab:5}).

So far, we assumed that the matter that obscures the AGN
is located only on the line of sight.  
If the absorbing matter covers a significant solid angle viewed 
 from the AGN such as in the case of a torus,
some of the incoming photons will be scattered into our line of sight.
Then the model given by eq.~\ref{eq:tra} would overestimate the
AGN luminosity.
Matt, Pompilio, \& Franca (1999) performed a Monte Carlo calculation  
of the X-ray transmission through spherically distributed 
matter for various column densities.
According to their result,
the true AGN luminosity would be smaller by a factor of $\sim2$,
if the absorbing matter is distributed in a spherical geometry.
Consequently, for the same flux from the cold reflector,
the true value of $R$ should be larger by a factor of 2 than that
in Table~5.
The column density of the absorbing matter is also subject
to a slight overestimation.

The case that a heavy absorber exists only on the line of sight is
rather unlikely. On the other hand, since the fitting result shows
a relatively small absorption for the reflection component,
the absorbing matter is probably not covering the entire sphere.
There two cases, a cloud on the line of sight are spherically
distributed matter, are considered to represent two extremes.

In conclusion, 
the {\it reflected- and transmitted-}AGN model gives
the following AGN parameters:
$\Gamma = 1.33-2.02$,
$L_{\rm X}$(2--10keV) = $5\times10^{43}$ -- $6\times10^{44}$ ergs/s,
$R$ = 0.1--1,
and $N_{\rm H,AGN}$ = 1.0 -- 2.7$\times10^{24}$ cm$^{-2}$,
where the range of parameters represents
not only the statistical errors but also
the uncertainties due to unknown inclination angle of the reflector,
and the limits for the two extreme geometries of 
the absorbing/scattering material.

\begin{table}
\label{tab:5}
\begin{center}
  \caption{Reflection and transmitted AGN model}
  \begin{tabular}{lll}
    \hline 
    \hline 
    Parameter              & Value (fit 1)          & Value (fit 2)$^{a)}$ \\
    \hline 
    $\Gamma$               & 1.59$^{+0.42}_{-0.26}$ & 1.71$^{+0.31}_{-0.37}$ \\
    $\mu = cos \theta$     & 0.45 (fix)      & $^{b)}$0.91$^{+0.04}_{-0.86}$ \\
    Fe abundance (solar)   & 0.85$^{+1.40}_{-0.65}$ & 0.77$^{+1.23}_{-0.47}$ \\
    $K_{\rm AGN}$(ASCA)$^{c)}$ & 9.1$\times10^{-3}$     & 1.3$\times10^{-2}$ \\
    $N_{\rm H,AGN}$ ($\times10^{24}$cm$^{-2}$)  
			   & 1.7$^{+1.0}_{-0.5}$    & 1.7$^{+0.7}_{-0.5}$ \\
    $R$               & 0.51$^{+0.99}_{-0.41}$ & 0.34$^{+0.66\ a)}_{-0.21}$ \\
    $K_l$(counts s$^{-1}$ cm$^{-2}$)$^{d)}$ 
	                   & 2.4$\times10^{-5}$ & 2.3$\times10^{-5}$ \\
    $E_l$(keV)             & 6.44$^{+0.04}_{-0.05}$ & 6.44$^{+0.04}_{-0.05}$ \\
    $K_{\rm AGN}(RXTE)$/$K_{\rm AGN}(ASCA)$  
                           & 1.00$^{+0.18}_{-0.15}$ & 1.03$^{+0.18}_{-0.16}$ \\
    $\chi^2/\nu$           & 86.2/85                & 84.8/84  \\
    \hline 
  \end{tabular}
\end{center}
\begin{itemize}
  \item[$^{a)}$]: In the fit, $\mu$ is left free. Maximum value of $R$
	is assumed to be 1.
  \item[$^{b)}$]: The lower and upper limits correspond to
              the limitation of the model code.
  \item[$^{c)}$]: In the unit of counts s$^{-1}$ cm$^{-2}$ keV$^{-1}$ @1keV.
  \item[$^{d)}$]: The physical line width is assumed to be zero.
 	The equivalent width with respect to the reflection
	continuum is 0.96~keV and 0.95~keV for the fit 1 and 2, respectively.
\end{itemize}
\end{table}

\section{Discussion}
The X-ray spectrum of NGC~6240 is satisfactorily explained with a model
consisting of a thermal component and an AGN component.
For the AGN component, and equally good fit is obtained either by the
reflection-only model or by the {\it reflected- and transmitted-}AGN model.
Hence, we cannot conclude the detection of the direct AGN component from
the present data. However, without the direct (transmitted) component
the derived photon index is noticeably smaller than the typical
values for AGN. We consider it more probable that the direct AGN component
is present at high energies.

According to the unified model for Seyfert galaxies
(e.g. Antonucci 1993), 
one can interpret the result in terms of a tilted molecular torus 
in which the near-side of the torus acts as an absorber 
and the far-side acts as a reflector.
If the torus is axisymmetric,
both the reflecting part and the absorbing part would have
a similar column density.
With the {\it reflected- and transmitted-}AGN model,
the column density of the obscuring matter is estimated
to be $\sim2\times10^{24}$ cm$^{-2}$, corresponding to
Thomson optical depth of $\sim1$.
This is practically enough for a reflector.
Thus, the present result is consistent with the interpretation
that NGC~6240 has a similar geometry to Seyfert 2 galaxies.
The obtained column density gives an IR extinction of
$A_{K}\sim100$, which is consistent with the fact
that the previous IR observations did not find the AGN itself.
Narrow line region that has not been seen in the optical observations
would be obscured as well.

The intrinsic X-ray luminosity of NGC~6240 is estimated in \S~4.3
to be in the range $5\times10^{43}-6\times10^{44}$ ergs/s in the range 
2--10~keV. For the reflection-only case (\S~4.2), 
the luminosity is estimated to be larger than $4\times10^{43}$ ergs/s,
while the upper bound of luminosity is not determined. 
However, it would be plausible that
the solid angle factor $R$ is not much less than several \%, which gives
essentially the same luminosity range as the above.
This luminosity is among those of the most luminous Seyfert nuclei, and
even comparable to those of quasars.
The bolometric luminosity of the AGN may well exceed $10^{45}$ erg/s.

Concerning the power source of the huge IR luminosity,
a measure of the contribution of an AGN to the IR emission
is a ratio of X-ray luminosity to IR luminosity,
$L_{\rm X}/L_{\rm IR}$.
Based on our results given above and the IR flux calculated
 from the IRAS result with the formula,
$F_{\rm IR}$ = flux(25$\mu$m)$\times$($\nu_{25\mu{\rm m}}$)
+ flux(60$\mu$m)$\times$($\nu_{60\mu{\rm m}}$),
gives $L_{\rm X}{\rm(2-10~keV)}/L_{\rm IR} = 0.01-0.1$ for NGC~6240.
This agrees with those of other Seyfert nuclei given by
Ward et al. (1988).

Vignati et al. (1999), using the {\it BeppoSAX} data of NGC~6240,
claim that the direct AGN component was positively detected, 
though they assumed the photon index of the AGN to be 1.8. 
Other than that, their results and interpretation are essentially
in agreement with ours.

The luminosity of the AGN can be used to infer
the mass of the central black hole in NGC~6240.
If the AGN accretes about 1\% of the Eddington luminosity,
the total luminosity of $10^{45}$ erg/s would lead to
a mass of $M_{\rm AGN} \simeq 10^9 M_{\odot}$,
which is consistent with the value expected from
the galaxy-mass to black-hole-mass relation (Lauer et al. 1997),
with the estimated galaxy mass of $10^{11-12} M_{\odot}$ 
after having completed its merging epoch (Shier \& Fisher 1997).

\section{Conclusion}
The 0.5--200~keV wide band energy spectrum of NGC~6240
obtained with {\it RXTE} and {\it ASCA} is accounted for in terms of
a soft thermal component and a hard AGN component.
The soft component is presumably due to star burst activity.
The AGN component consists of a Compton reflection component
accompanied by an intense Fe-K emission line and probably
a transmitted component
(a direct component penetrating through a thick absorber).
The detection of the transmitted component is not conclusive from the
fitting.
However, without a transmitted component, the photon index is unusually small.
Assuming that the solid angle factor $R$ does not exceed 1,
we estimated the intrinsic X-ray luminosity of the AGN 
in the range 2--10~keV to be in the range $4 \times 10^{43}-6\times10^{44}$
ergs/s, which yields the ratio of the X-ray luminosity (2--10~keV) to the
IR luminosity of $0.01-0.1$.
The column density that obscures the central AGN is estimated
to be larger than $1.0\times 10^{24}$ cm$^{-2}$.
These results show that NGC~6240 is among the most luminous Seyfert 2 
galaxies.

\section*{Acknowledgments}
The authors are grateful to Dr. Joachim Siebert 
for helping the {\it RXTE} data analysis.
Y.I. was supported by the post-doctoral program of the 
Max-Planck-Gesellschaft and is currently supported by the
Japan Society for the Promotion of Science Postdoctoral Fellowships
for Research Abroad.
KML gratefully acknowledges support by NAG5-6921 (RXTE)
and NAG5-7971 (LTSA).

\label{lastpage}
\end{document}